\documentclass[smallextended]{svjour3}       
\usepackage{graphicx}
\usepackage{gensymb}
\usepackage{upgreek}
\usepackage{lineno}
\usepackage{url}
\usepackage[colorlinks=true,linkcolor=blue,citecolor=blue,urlcolor=blue,]{hyperref}
\usepackage{doi}
\usepackage{cite}
\usepackage[misc]{ifsym}
\journalname{Experimental Astronomy}

\begin{document}

\title{Design and test of a portable Gamma-Ray Burst simulator for GECAM}


\author{Can Chen \textsuperscript{1,2}
\and Shuo Xiao \textsuperscript{1,2}
\and Shaolin Xiong \textsuperscript{1}
\and Nian Yu \textsuperscript{1,2}
\and Xiangyang Wen \textsuperscript{1}
\and Ke Gong \textsuperscript{1}
\and Xinqiao Li \textsuperscript{1}
\and Chaoyang Li \textsuperscript{1,3}
\and Dongjie Hou \textsuperscript{1}
\and Xiongtao Yang \textsuperscript{1}
\and Zijian Zhao \textsuperscript{1}
\and Yuxuan Zhu \textsuperscript{1, 4}
\and Dali Zhang \textsuperscript{1}
\and Zhenghua An \textsuperscript{1}
\and Xiaoyun Zhao \textsuperscript{1}
\and Yupeng Xu \textsuperscript{1,2,*}
\and Yusa Wang \textsuperscript{1,*}
}

\institute{
\Letter Yupeng Xu\newline
\email{xuyp@ihep.ac.cn}\newline
\Letter Yusa Wang\newline
\email{wangyusa@ihep.ac.cn}\newline
    \at{1} Key Laboratory of Particle Astrophysics, Institute of High Energy Physics, Chinese Academy of Sciences, Beijing 100049, China.
    \at{2} University of Chinese Academy of Sciences, Chinese Academy of Sciences, Beijing 100049, China.
    \at{3} Physics and space science college, China West Normal University, Sichuan, 637002, China
    \at{4} College of Physics, Jilin University, Changchun 130012, China
}
\date{Received: date / Accepted: date}

\maketitle

\begin{abstract}
The main scientific goal of the Gravitational wave high-energy Electromagnetic Counterpart All-sky Monitor (GECAM) is to monitor various types of Gamma-Ray Bursts (GRB) originated from merger of binary compact stars, which could also produce gravitational wave, and collapse of massive stars. In order to study the response of GECAM Gamma-Ray Detectors (GRDs) to high-energy bursts and test the in-flight trigger and localization software of GECAM before the launch, a portable GRB simulator device is designed and implemented based on grid controlled X-ray tube (GCXT) and direct digital synthesis (DDS) technologies. The design of this GRB simulator which modulates X-ray flux powered by high voltage up to 20 kV is demonstrated, and the time jitter (FWHM) of the device is about 0.9 $\upmu$s. Before the launch in December, 2020, both two GECAM satellites were irradiated by different types of GRBs (including short and long bursts in duration) generated by this GRB simulator.  
The light curves detected with GECAM/GRDs are consistent with the programmed input functions within statistical uncertainties, indicating the good performance of both the GRDs and the GRB simulator.

\end{abstract}


\section{Introduction}\label{Introduction}
Gamma-Ray Burst (GRB), the most violent explosion after the Big Bang \cite{Zi-GaoDai2018GRBs?}, is a phenomenon that gamma-rays from the distant universe suddenly increase in a short time scale. 
GRBs are usually divided into long bursts or short bursts according to the duration of $\rm T_{90}$ which is the time interval between 5\% and 95\% of the cumulative observed fluence above the background \cite{kouveliotou1993identification}. Short bursts are believed to originate from binary compact star mergers, which has been firmly demonstrated by the first binary neutron star merger (GW 170817) and its high-energy electromagnetic counterpart (GRB 170817A) \cite{abbott2017gravitational}\cite{goldstein2017ordinary}, while long bursts are produced by collapse explosions of massive stars.

\begin{figure}[htb]
	\begin{center}
		\includegraphics[width=0.3\textwidth]{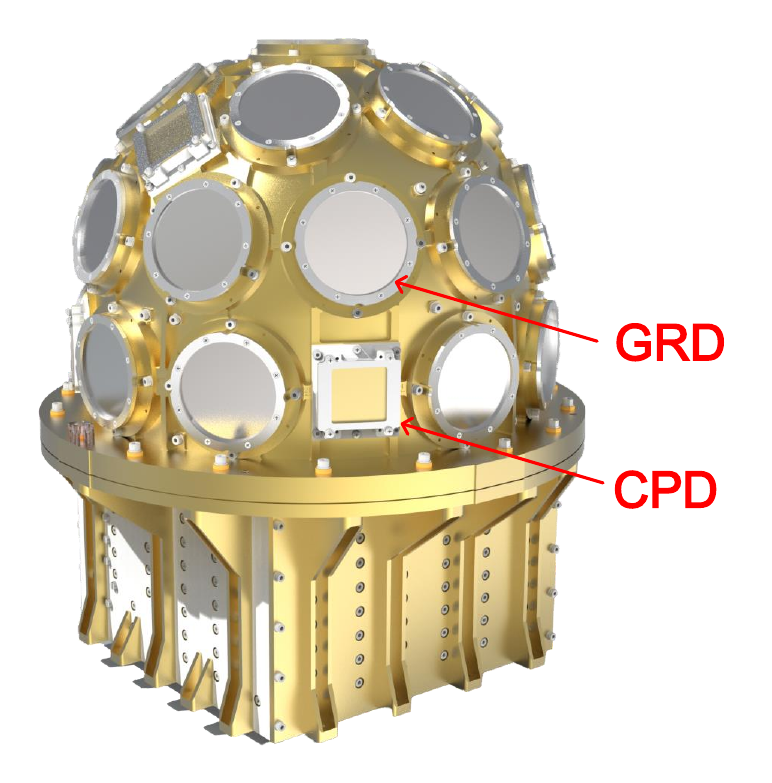}
		\includegraphics[width=0.3\textwidth]{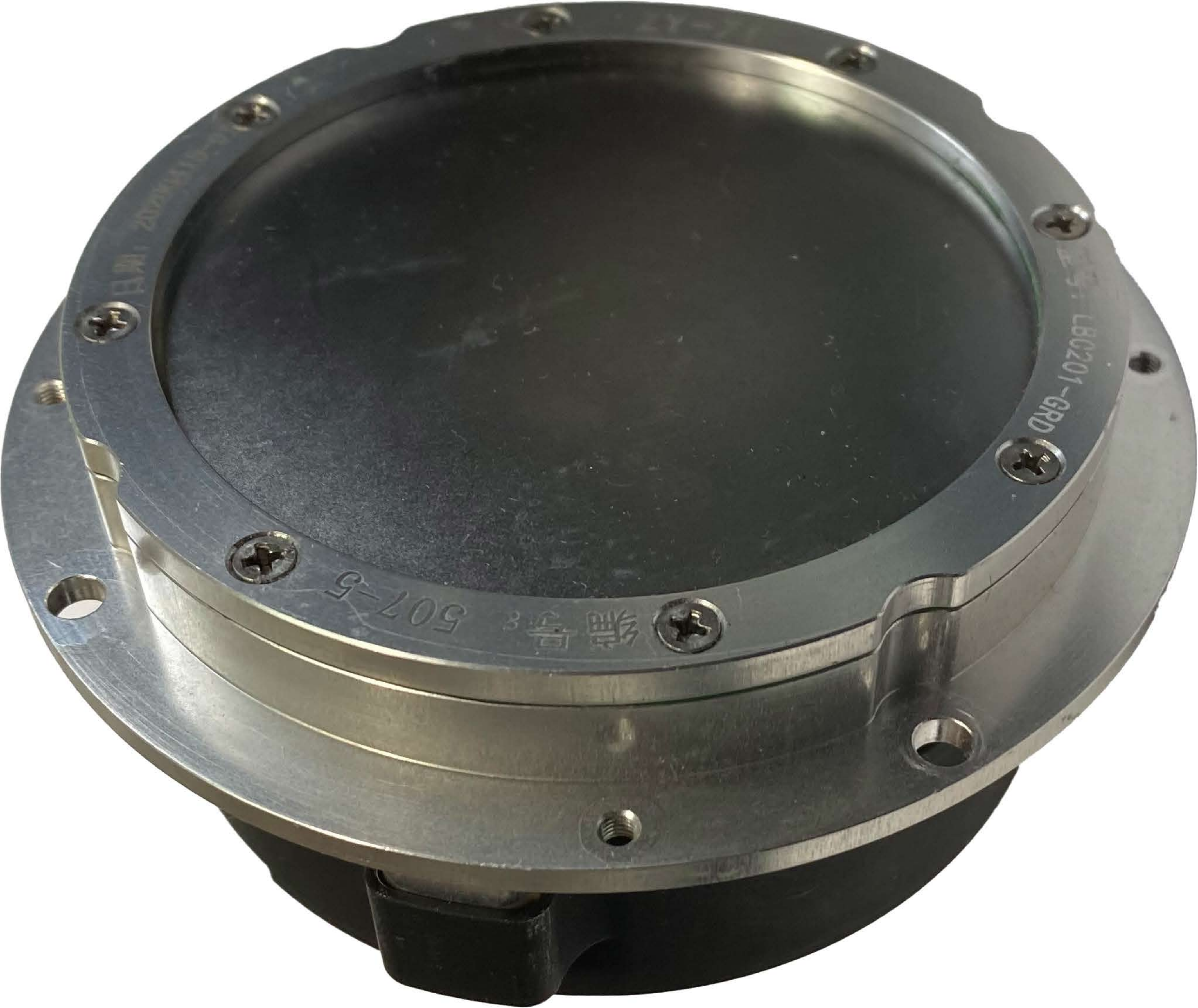}
		\caption{Left: Each GECAM satellite is equipped with 25 GRDs and 8 CPDs. Right: The photo of a GRD of GECAM.
		}
		\label{GRD}
	\end{center}
\end{figure}

To monitor the high-energy electromagnetic radiation of gravitational waves and other gamma-ray bursts, Gravitational wave high-energy Electromagnetic Counterpart All-sky Monitor (GECAM) was proposed in 2016. GECAM is composed of two microsatellites\cite{xiong50special}, and each GECAM satellite contains 25 Gamma-Ray Detectors (GRDs) and 8 Charged Particle Detectors (CPDs). The illustration of a GECAM satellite is shown in Fig.~\ref{GRD} (left). As the main scientific payload, GRD is implemented by coupling $\rm LaBr_3$:Ce with the Silicon Photomultiplier (SiPM) array, and the geometry area of each GRD is about 45 $\rm cm^2$. The photo of GRD is shown in Fig.~\ref{GRD} (right). 
Each GRD is equipped with a high gain and a low gain signal processing channel \cite{zhang2019energy}\cite{lv2018low} to achieve a wide dynamic detection range (about 6 to 5000 keV).

Response of detectors to different type of GRBs could be studied with numerical simulations based Monte Carlo modeling, as for example, it has been done for the Gamma-Ray Burst Monitor (GBM), one of the scientific payloads of the Fermi \cite{bissaldi2009ground}. 
However, to study the real behavior of GECAM/GRD to GRBs and verify the in-flight trigger and localization software (X.Y. Zhao et al. In preparation), an X-ray source with programmable light curves and intensities to mimic various gamma-ray bursts on-ground is required. 

Stroboscopic X-ray generators based on condenser-discharge induced flash X-ray technique were designed and used for X-ray imaging in 1999 \cite{sato1999condenser}, however, this method cannot produce X-rays with continuous varying photon intensity.
Pulse X-ray source based on chopper was designed in 2011 \cite{hui2011simulation}, but its turntables need to be replaced if one wants to produce a different light curve, which leads to poor flexibility. 

In this study, the GRB simulator was designed based on the grid controlled X-ray tube (GCXT) which had been used for X-ray communication and X-ray pulsar-based navigation simulation source \cite{LizhiSheng2014}\cite{zhou2013analog}. 
Based on the traditional X-ray tube, a grid electrode is added between the anode and the cathode to control the flux of electrons bombarding the anode by applying a programmable negative voltage vs the cathode. 
Thus, a GCXT can generate high-energy photons with variable intensity. 
As in a triode system, the grid cut-off voltage determining the state of negligible tube current is directly related to the anode voltage and so to the maximum X-ray energy. 
Due to the modulation voltage swing, the energy modulation range is limited to the soft X-ray band (about 1 to 10 keV) \cite{LizhiSheng2013x}, which is relatively lower than the observation energy range (about 6 keV to 5 MeV) of GECAM \cite{li2020gecam}. 
A newly released high-voltage precision operational amplifier ADHV4702-1 manufactured by Analog Devices Inc. \cite{ADHV4702} is applied in the GRB simulator to achieve the modulation of X-ray at higher energies (up to about 20 keV). 
In addition, considering the complexity and irregularity of actual GRB light curve, a method of directly synthesizing arbitrary waveforms called direct digital synthesis (DDS)\cite{Vankka1997} is adopted in making a modulation voltage signal source.

This paper is structured as follows. The design of the GRB simulator is introduced in Section~\ref{Design}. Next, the performance of the GRB simulator is presented in Section~\ref{Performance}. Then, the GRB ground tests with the GRB simulator performed before the launch of GECAM is depicted in Section~\ref{experimen}. Finally, the performance of the GRB simulator and the response of GRDs to GRBs are summarized in Section~\ref{Conclusion}.

\section{Design of the GRB simulator}\label{Design}
\begin{figure}[htb]
	\centering
	\includegraphics[width=0.7\textwidth]{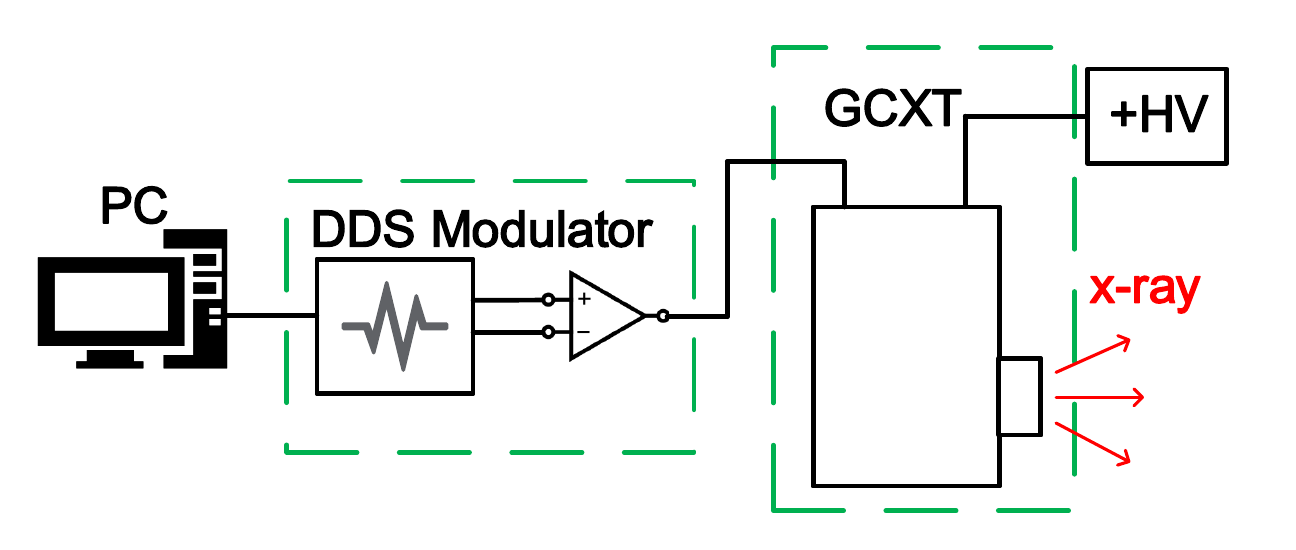}
	\caption{Schematic of the GRB simulator mainly consist of DDS modulator and GCXT.}
	\label{grb_simulator}
\end{figure}

The GRB simulator is mainly composed of two parts: DDS modulator and GCXT. Each part is modularized to make the device portable and convenient to upgrade, as shown in Fig.~\ref{grb_simulator}. 
The firmware containing GRB light curve data is loaded into the Field-Programmable Gate Array (FPGA) inside the DDS modulator through the computer. 
Then the DDS modulator generates a modulation signal transmitted to the grid electrode of GCXT. 
Finally, GCXT produces the corresponding X-ray flux following the input light curve.

\subsection{DDS modulator}\label{DDS waveform generator}

DDS, a technique of directly synthesizing analog waveform, is utilized to generates artificial GRBs with arbitrary light curves. 
The DDS modulator is mainly realized with three evaluation boards, as shown in Fig. 3, a controller based on Altera EP4CE6E22C8 FPGA, a Digital-to-Analog Converter (DAC) board consist of an ADI 14-bit AD9767 DAC with parallel interface to the FPGA, and a driver module with a high voltage op-amp ADHV4702-1.

In order to accurately reproduce a light curve with timescale of the shortest GRBs (about 2 ms) \cite{golkhou2015energy}, a stable global clock with frequency of $f_{\rm clk} = 100 \, {\rm MHz}$ inside the FPGA is generated with PLL (Phase Locked Loop) frequency multiplication through the on-board $\rm 50\, MHz$ crystal oscillator.
The output frequency of a light curve phase point is managed with a $N$-bit phase accumulator register and a frequency control word register with the value of $K$ (ranging from 1 to $2^N - 1$). 
The phase accumulator accumulates continuously with the addition of $K$ at frequency $f_{\rm clk}$. 
A phase point is determined and the accumulator resets to zero when it overflows. 
Thus, the output frequency is the rate of the phase accumulator overflows \cite{Vankka1997}:
   \begin{equation}\label{phase accumulator clock}
      f_{\rm out} = f_{\rm clk} \cdot K/2^N
   \end{equation}
where the value of $N$ determines the lowest phase clock could be obtained and is fixed to 32 in this study. 

The data describing the GRB light curve, stored in the look up table (LUT), are sent to the DAC through the parallel interface at each phase point, i.e. with frequencies of $f_{\rm out}$. Therefore, a nearly continuously adjustable output frequency could be controlled by the single parameter $K$ and the duration of the light curve, i.e. the number of phase points, is limited by the memory size of the LUT.

To drive the GCXT, a high voltage amplifier is essential. Considering the future requirement to achieve higher energy X-ray modulation (at least 50 keV) and extendibility, ADHV4702-1 is used to amplify the modulation signal from the DAC. ADHV4702-1 is the first 220 V high-voltage precision operational amplifier with a wide dynamic range (-110 V$\sim$+110 V) and a relatively high slew rate (74V/µs typical) \cite{ADHV4702}.

\begin{figure}[htb]
	\begin{center}
		\includegraphics[width=1\textwidth]{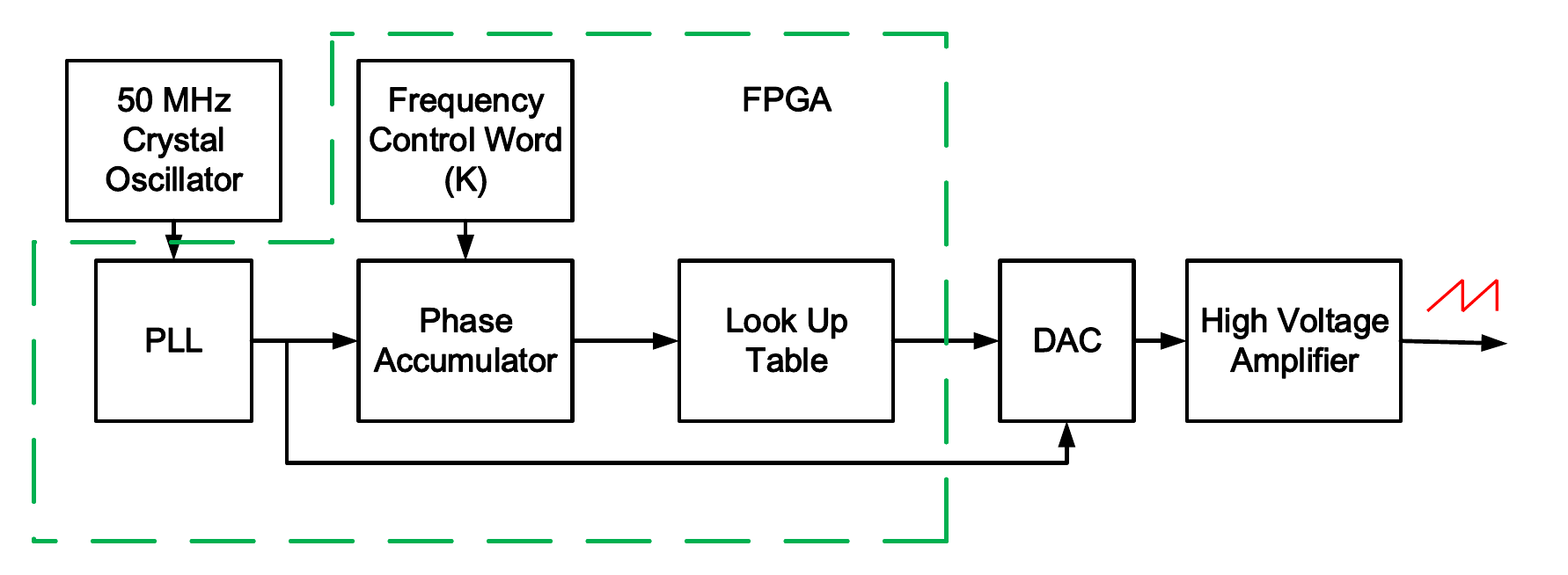}
		\includegraphics[width=0.4\textwidth]{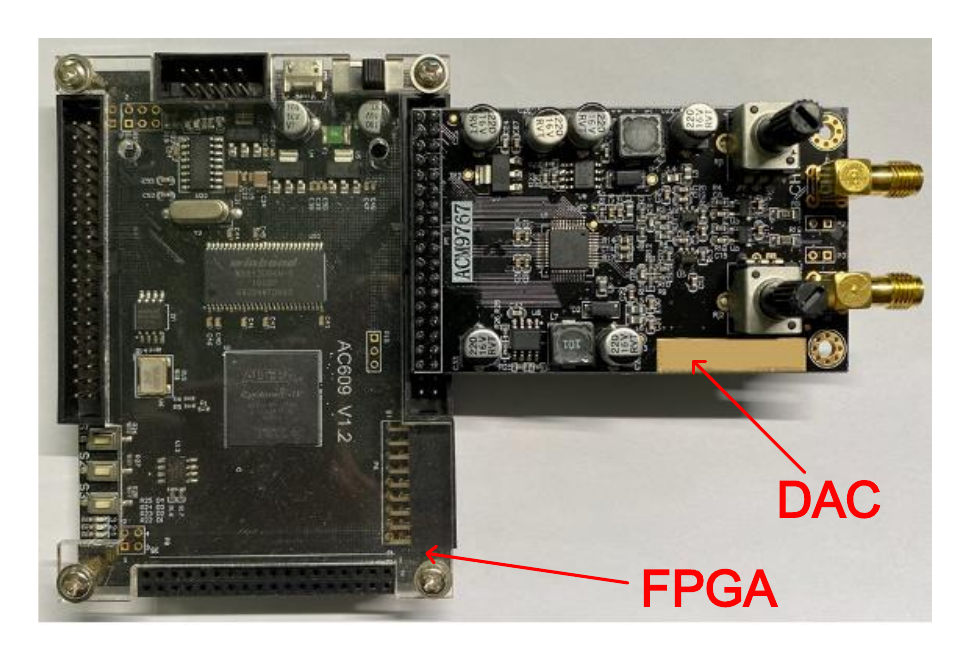}
		\includegraphics[width=0.36\textwidth]{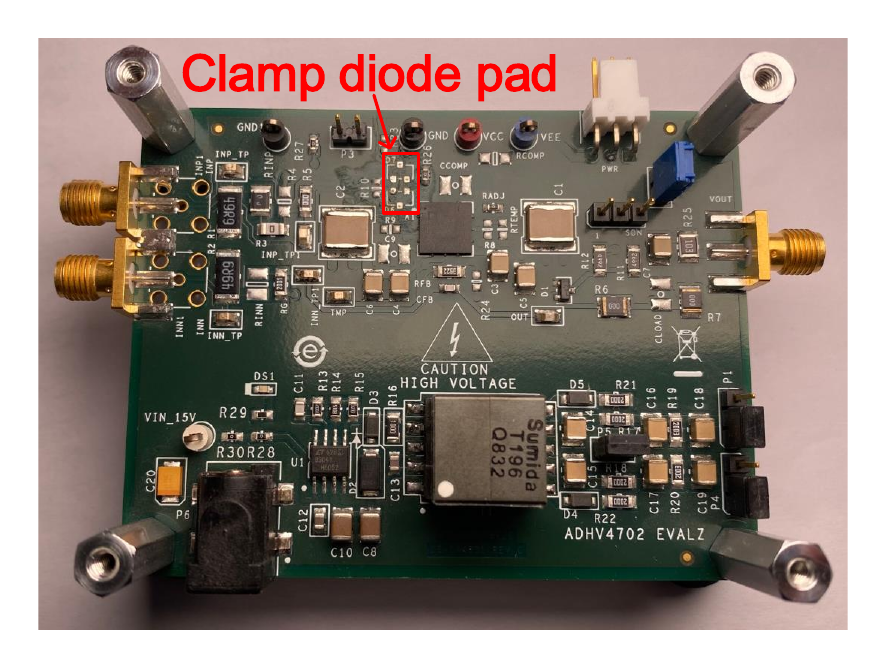}
		\caption{Upper: Simplified schematic of DDS modulator. Lower-left: Circuit board of FPGA and DAC. Lower-right: ADI ADHV4702-1 high-voltage precision operational amplifier board whose clamp diode is removed to maximizes slew rate.
		}
		\label{DDS}
	\end{center}
\end{figure}

\begin{figure}[htb]
	\begin{center}
		\includegraphics[width=0.5\textwidth]{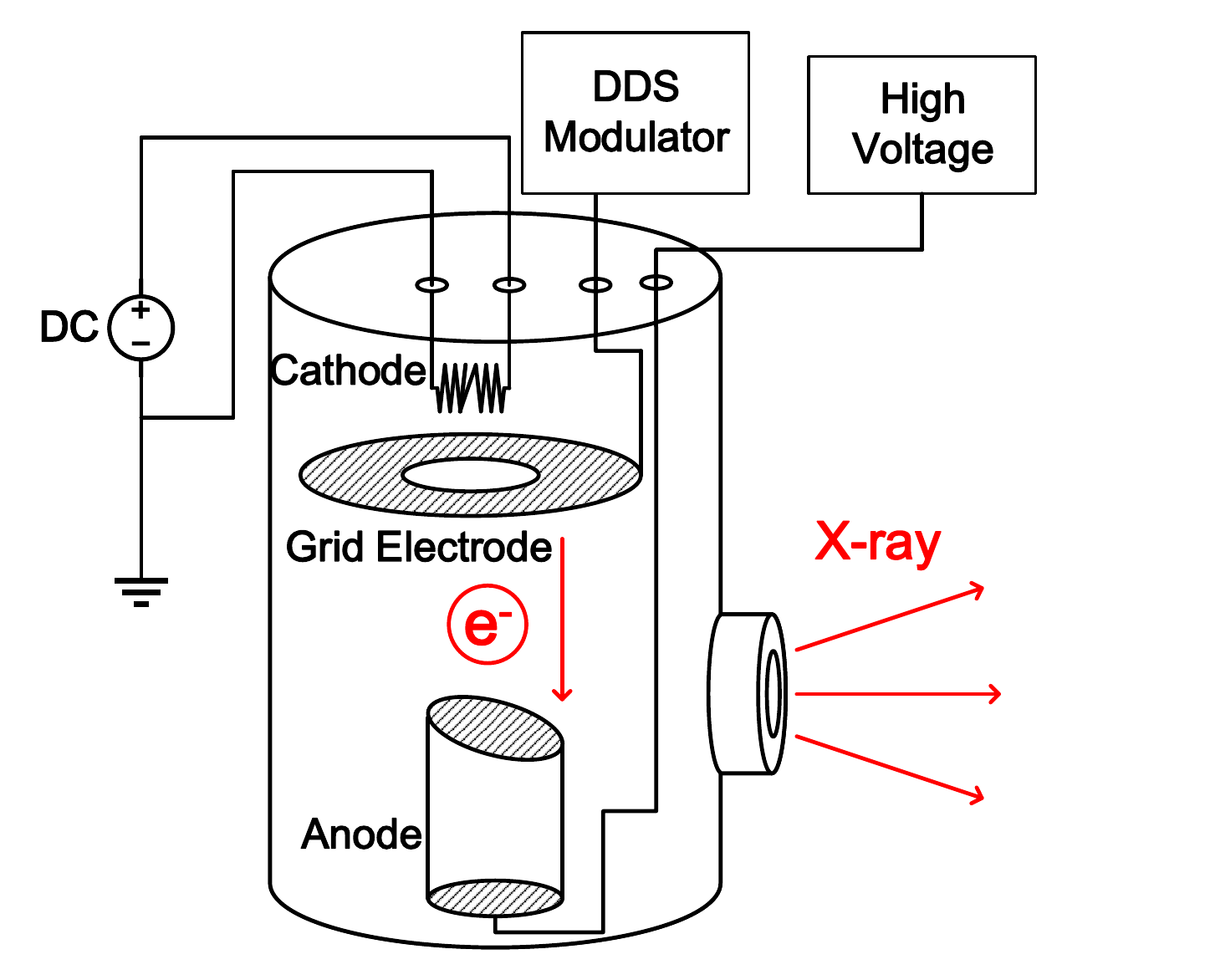}
		\caption{Simplified schematic of the GCXT, mainly composed of cathode, grid electrode and anode.}
		\label{GCXT_sch}
	\end{center}
\end{figure}

\subsection{Design of grid controlled X-ray tube}
\label{Design of grid controlled X-ray tube}
The GCXT we designed is in a cylindrical shape with a height of 17 cm and a diameter of 8 cm. The simplified schematic of GCXT is shown in Fig.~\ref{GCXT_sch}. Based on the classical X-ray tube, a grid electrode with programmable negative voltage is added between the cathode and the anode. 
The tungsten cathode filament is heated by a controllable current source to emit thermal electrons with intensity proportional to the filament current. 
The electric field produced by the high voltage applied to the anode accelerates the thermal-emitted electrons that bombarding the anode target generates X-rays by bremsstrahlung including the characteristic lines of the anode material. 
Due to the thermal inertia, the output intensity can not be controlled by the filament current for fast timing signal simulation. 
Instead, a negative voltage applied to the grid electrode close to the cathode could control the electron flux passing through promptly, and therefore, control the intensity of the X-ray generated by the X-ray tube.

\section{Performance of the GRB simulator}\label{Performance}

\subsection{Grid voltage modulation properties of the GRB simulator}\label{GCXT modulation properties}
\begin{figure}[htb]
	\begin{center}
		\includegraphics[width=0.45\textwidth]{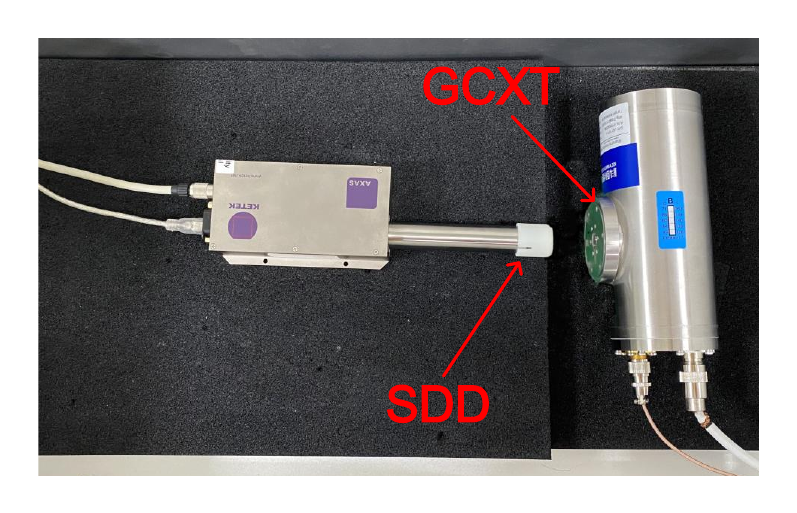}
		\includegraphics[width=0.45\textwidth]{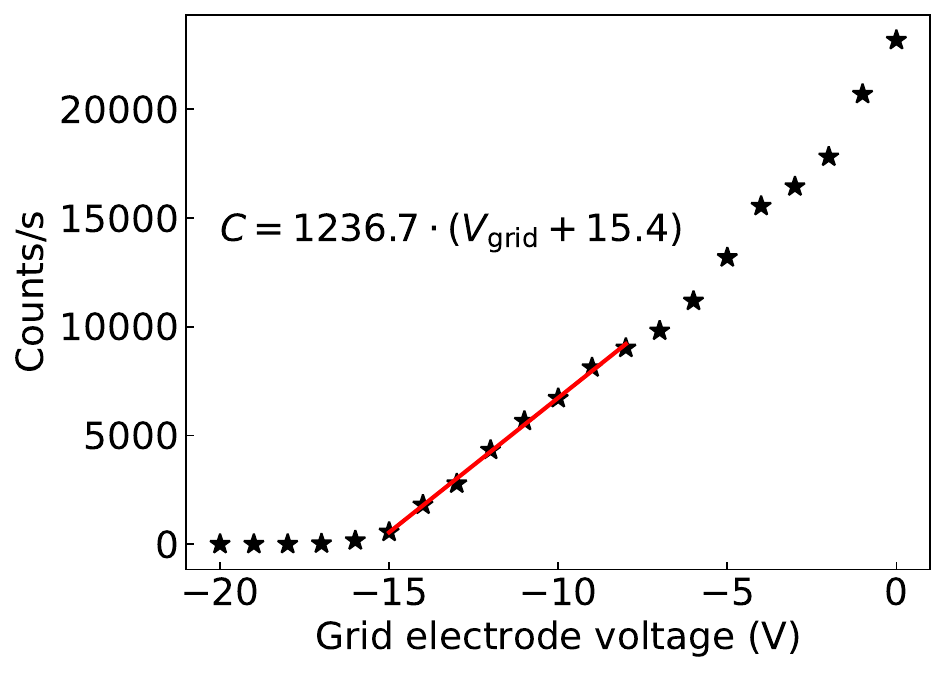}
		\caption{Left: Experimental layout to investigate the modulation properties of GCXT. Right: Output count rates of GCXT versus grid electrode voltage.}
		\label{GCXT_expr}
	\end{center}
\end{figure}

To verify the performance of modulating X-ray flux generated by the GCXT, a KETEK AXAS-D silicon drift detector (SDD) is used to measure X-ray flux and spectrum emitted from the GCXT under different grid voltages. 
The experimental layout is shown in Fig.~\ref{GCXT_expr} (left). The SDD with 30 $\rm mm^2$ active area is placed about 5 cm from the GCXT. The anode voltage is set to +20 kV supplied by the Wisman High voltage source XEL20P, while the voltage of the cathode is fixed to 0 V. 

\begin{figure}[htb]
	\centering
	\includegraphics[width=0.5\textwidth]{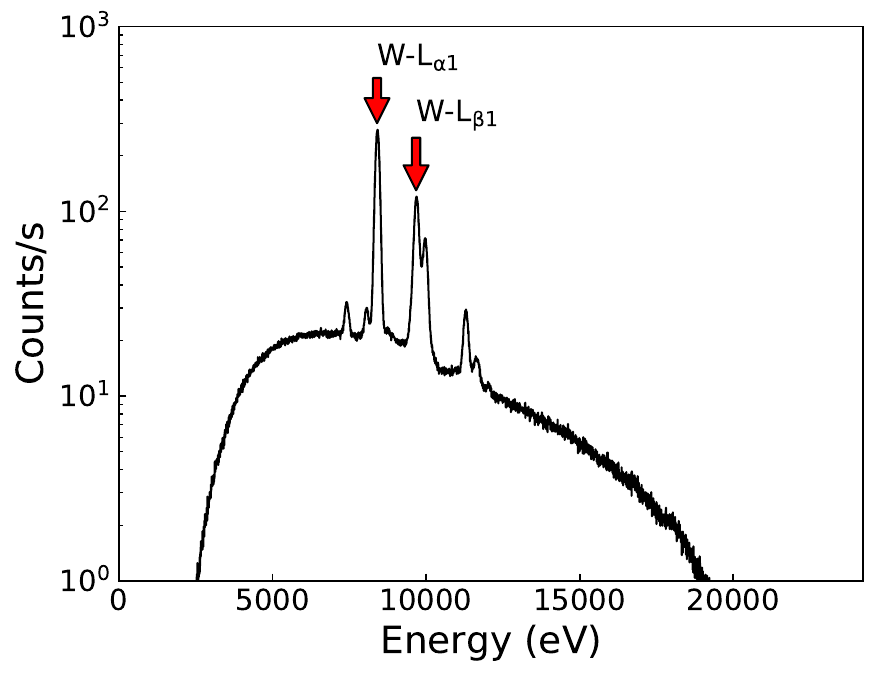}
	\caption{The raw energy distribution of GCXT detected by KETEK SDD when the high voltage source is set to 20 kV. The characteristic emission line of tungsten can be clearly seen, including the 8.40 keV W-$\rm L_{\upalpha1}$ line and the 9.67 keV W-$\rm L_{\upbeta1}$ line.}
	\label{Energy_spectrum_GCXT}
\end{figure}

The detected raw energy distribution is shown in Fig.~\ref{Energy_spectrum_GCXT}. 
Grid electrode voltage is set from 0 V to -20 V with a step size of 1 V to investigate the voltage response of GCXT, and the recording time of each voltage point is 100 s. The relationship between the count rates detected by SDD and the grid electrode voltage is shown in Fig.~\ref{GCXT_expr} (right). The upper threshold of grid electrode voltage to detect X-ray is about -18 V. We find that there is a relatively linear relationship between grid electrode voltage and output count rates in the range of -15 V to -8 V. The least squares method is adopted to testify the relationship between two variables:
    \begin{equation}
        C = 1236.7 \cdot (V_{\rm grid} + 15.4) 
        \label{equation_C_V}
    \end{equation}
where $C$ is the count rates (counts per second) acquired by SDD, $V_{\rm grid}$ is the voltage of grid electrode. 
Meanwhile, photon count rates increase with filament current, while maintaining the linear relationship in this voltage range. 
The GRB simulator works in this linear region during the following experiment to GECAM.

Thus, a specific GRB light curve can be generated by setting the corresponding grid electrode voltage at each phase point of a light curve.

\subsection{Time characteristic of the GRB simulator}
\label{GRB simulator time characteristic}
Time delay and time jitter between the application of the grid signals and the X-ray produced are caused mainly during the process of electron drifting and bombarding the anode, which directly affects the quality of GRB reproduced. 
Therefore, the time response of the GRB simulator is also investigated in this study.

\begin{figure}[htb]
	\centering
	\includegraphics[scale=0.4]{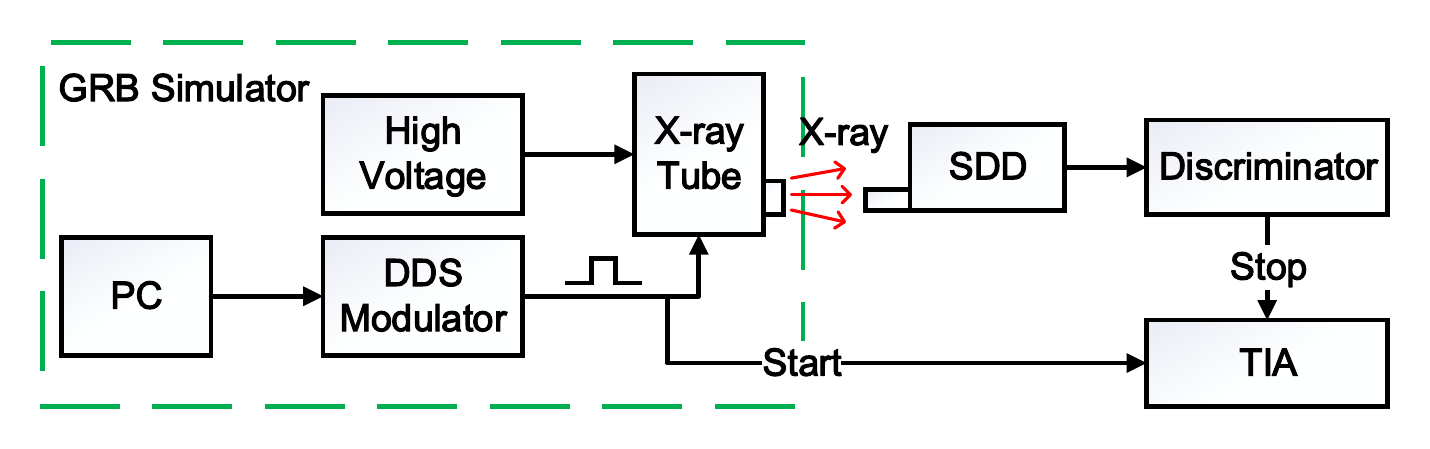}
	\caption{Simplified experimental schematic to investigate the time response of GCXT.}
	\label{time_experiment_GCXT}
\end{figure}

\begin{figure}[htb]
	\centering
	\includegraphics[scale=0.4]{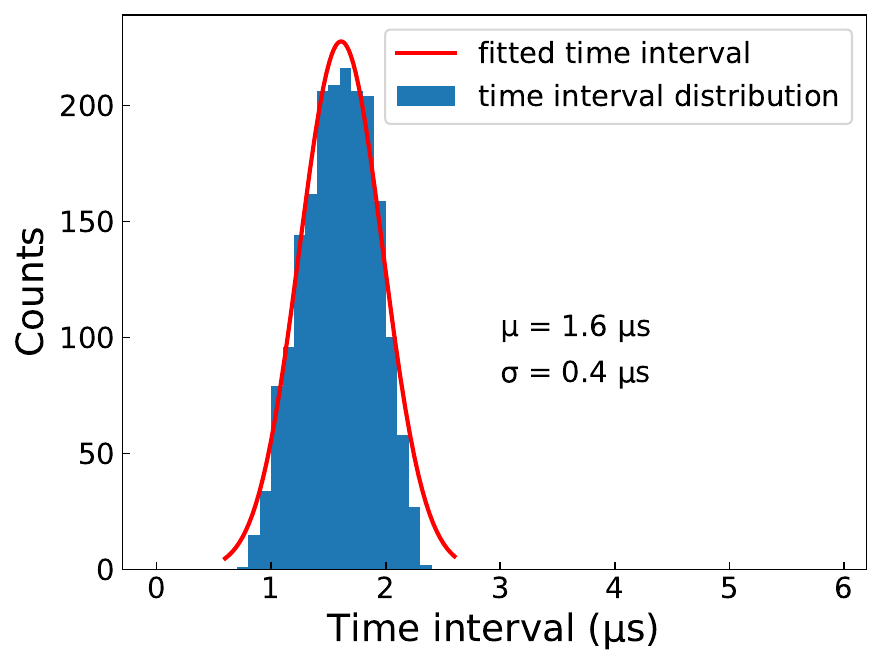}
	\caption{Time delay and time jitter of the GRB simulator. The blue line represents the detected time interval. The red line is the result of Gaussian fitting.}
	\label{time_characteristic_GCXT}
\end{figure}

The experimental schematic is shown in Fig.~\ref{time_experiment_GCXT}. The grid electrode of the GCXT is triggered by a square pulse signal with a duration of 1 $\upmu$s and a 100 kHz repeating frequency generated by the DDS modulator. 
Simultaneously, the same signal is sent to the Carmel NK732 time interval analyzer (TIA) as the start signal. NK732 is able to time tag events (edges of an input pulse train) at a rate of 20 million per second continuously with 2 ps resolution\cite{nk732}. 
The arrival time of the X-ray pulse indicated by the discriminator output of a homemade SDD shaping circuit\cite{yunian2020random} with a peaking time of about 0.55 $\upmu$s is used as the stop signal to the TIA. 
The distribution of time interval between the raising edge of the square pulse signal applied to the grid electrode (start signal) and arrival time of the first detected X-ray photon in the SDD (stop signal) is shown in Fig.~\ref{time_characteristic_GCXT}. 
The time jitter (Full Width at Half Maximum, FWHM) of the GRB simulator equals 0.9 $\upmu$s, and the time delay is 1.6 $\upmu$s with respect to the start of signal applied to the grid electrode.
Therefore, the minimum variable time scale of GRB light curve generated by the GRB simulator is 0.9 $\upmu$s. 
Taking into account the uncertain time introduced by the SDD signal shaping as well as the spread of photon generating time within square pulse, the time response of the GRB simulator measured here is a relatively rough estimate (upper limit).  

\section{GRB simulation experiment for GECAM on the ground }\label{experimen}
\begin{figure}[htb]
	\centering
	\includegraphics[scale=0.4]{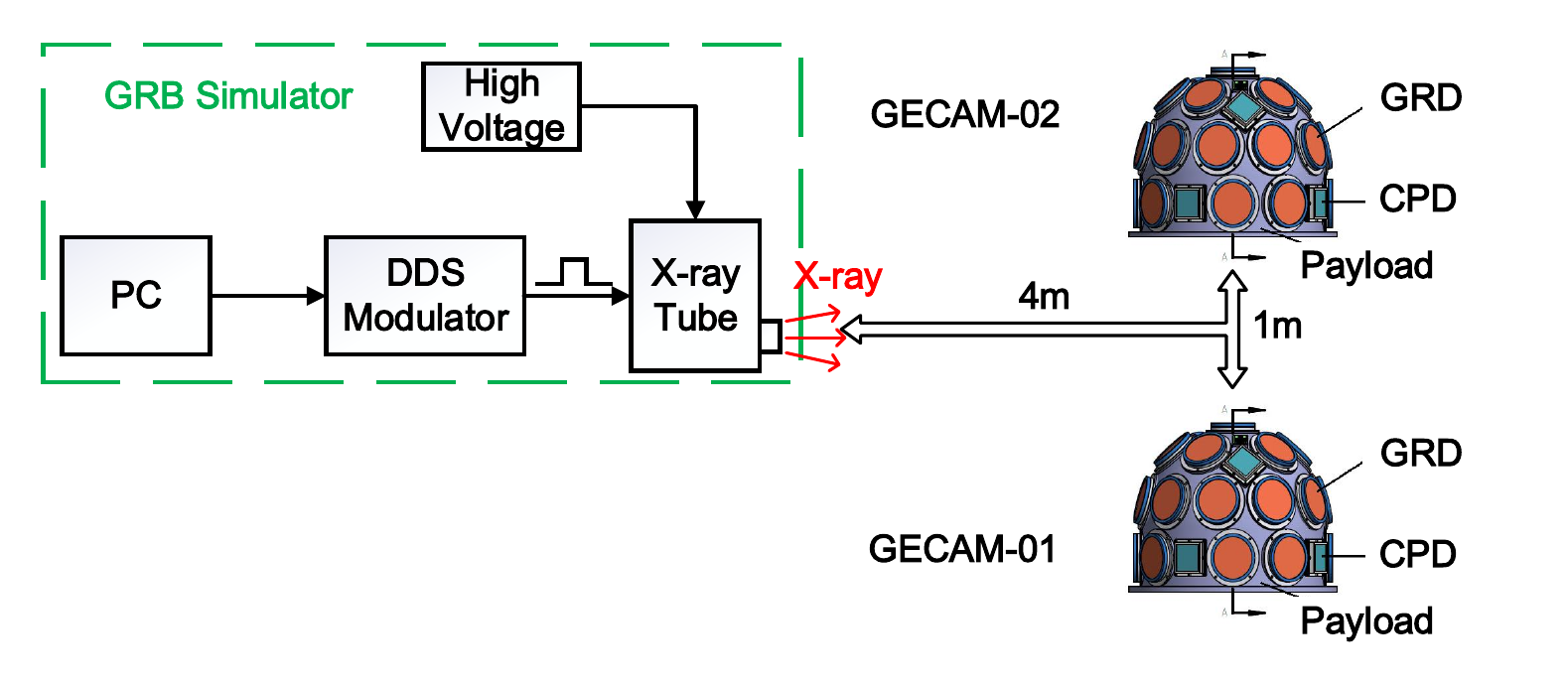}
	\caption{The layout of the ground GRB experimental to GECAM. GCXT is placed at the perpendicular bisector of GECAM-01 and GECAM-02 with about 4 m away from two satellites to allow X-rays to irradiating two satellites at the same time. The distance between the center of the two satellites is 1 m.}
	\label{GECAM_exper}
\end{figure}

Ground experiment to study the response to GRB and verify the in-flight trigger and localization software has been carried out to the two GECAM satellites (denoted as GECAM-01 and GECAM-02) comprehensively. The schematic diagram of this experimental layout is shown in Fig.~\ref{GECAM_exper}. Limited by the opening angle of GCXT photon output window (about 20\degree), the GCXT is placed at the perpendicular bisector of GECAM-01 and GECAM-02 with about 4 m away from these two satellites to allow X-rays irradiating both of them at the same time. 
The distance between the two satellites is about 1 m. 
Some GRDs (\#5, \#6, \#13, \#14, \#15, \#16, \#22, \#23, \#24)\cite{li2020gecam} facing to the X-ray tube can receive more X-ray photons because they have larger receiving area to the X-ray flux.
Considering the maximum photon energy generated by GCXT is about 20 keV, only the high gain channel of the GRD could detect these photons.

GRBs are commonly described as fast-rise exponential-decay (FRED) shapes (e.g. Fenimore et al. 1996)\cite{fenimore1996expanding}. 
However, the actual GRB light curve is complex and irregular. 
Hence, the shape of FRED and ``super-long'' curve synthesized with several actual GRBs that observed by {\it Fermi}/GBM were reproduced through the GRB simulator. 
In our tests, the duration of FRED shape GRB is set to 0.5 s, as a result, $N$ and $K$ should be set to 32 and 858993 respectively according to eq.~(\ref{phase accumulator clock}). 
The theoretical curve is the expected photon count rates detected by GRDs:
    \begin{equation}\label{theoretical_curve}
        C_{\rm theoretical}(t)=  a \cdot f (V_{\rm grid}(t))
    \end{equation}
where $a$ is a constant that relate to the response properties and effected area of different detectors, $f$ is the relationship between the grid electrode voltage and the output count rates of GCXT, as shown in eq. (\ref{equation_C_V}), $V_{\rm grid}(t)$ is the grid electrode voltage of each phase point in a light curve (i.e. the programmed input functions).

\begin{figure}[htb]
	\centering
	\includegraphics[scale=0.5]{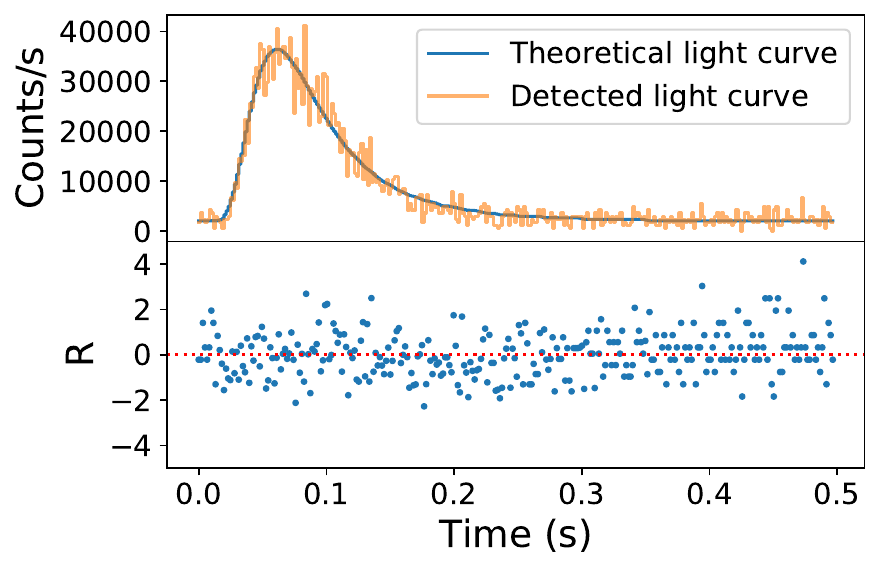}
	\includegraphics[scale=0.4]{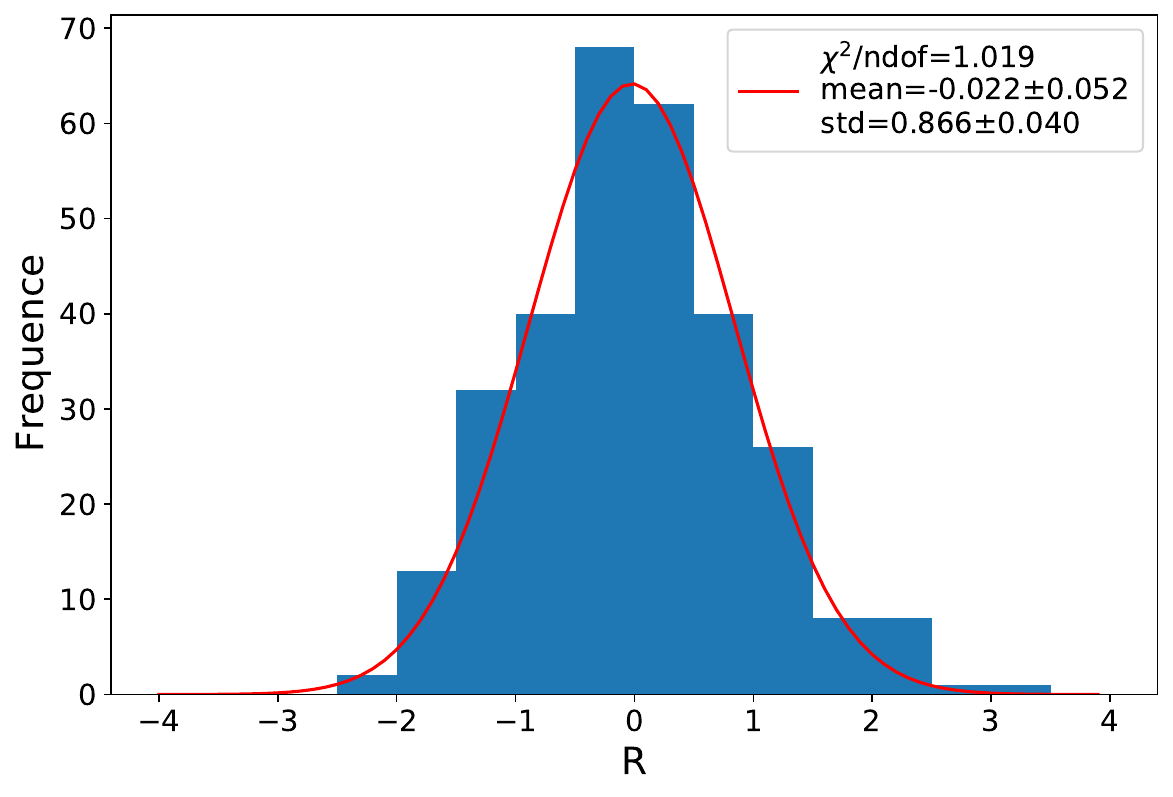}
	\caption{The upper panel is the FRED shape GRB light curve we constructed (blue line), then reproduced by the GRB simulator and detected by GRDs (orange line). 
	GRD\#5, \#6, \#13, \#14, \#15, \#16, \#22, \#23 and \#24 were used. 
	The bottom panel is the residual of the theoretical curve and the detected light curve.
	Bottom: The distribution of the residual. The red line is the result of Gaussian fitting.
	}
	\label{A-fred}
\end{figure}

\begin{figure}[htb]
	\centering
	\includegraphics[scale=0.5]{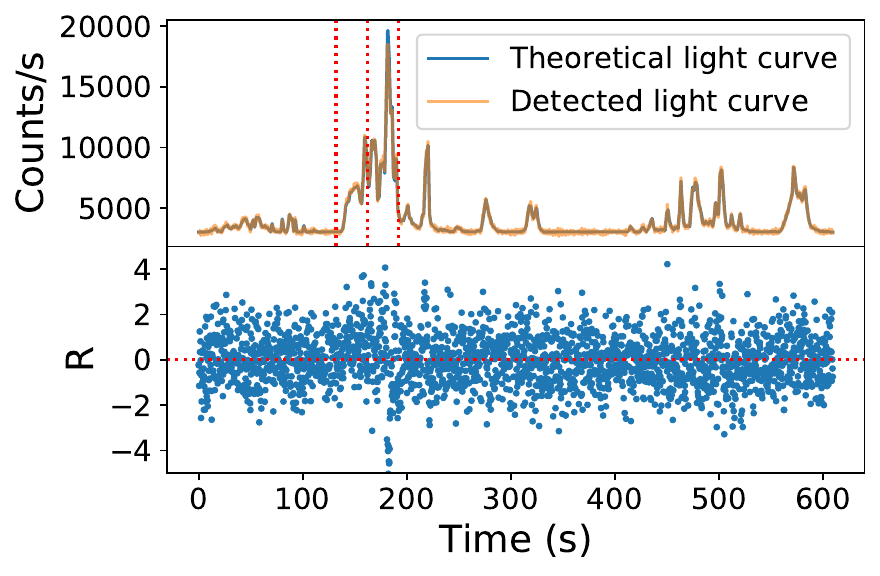}
	\includegraphics[scale=0.4]{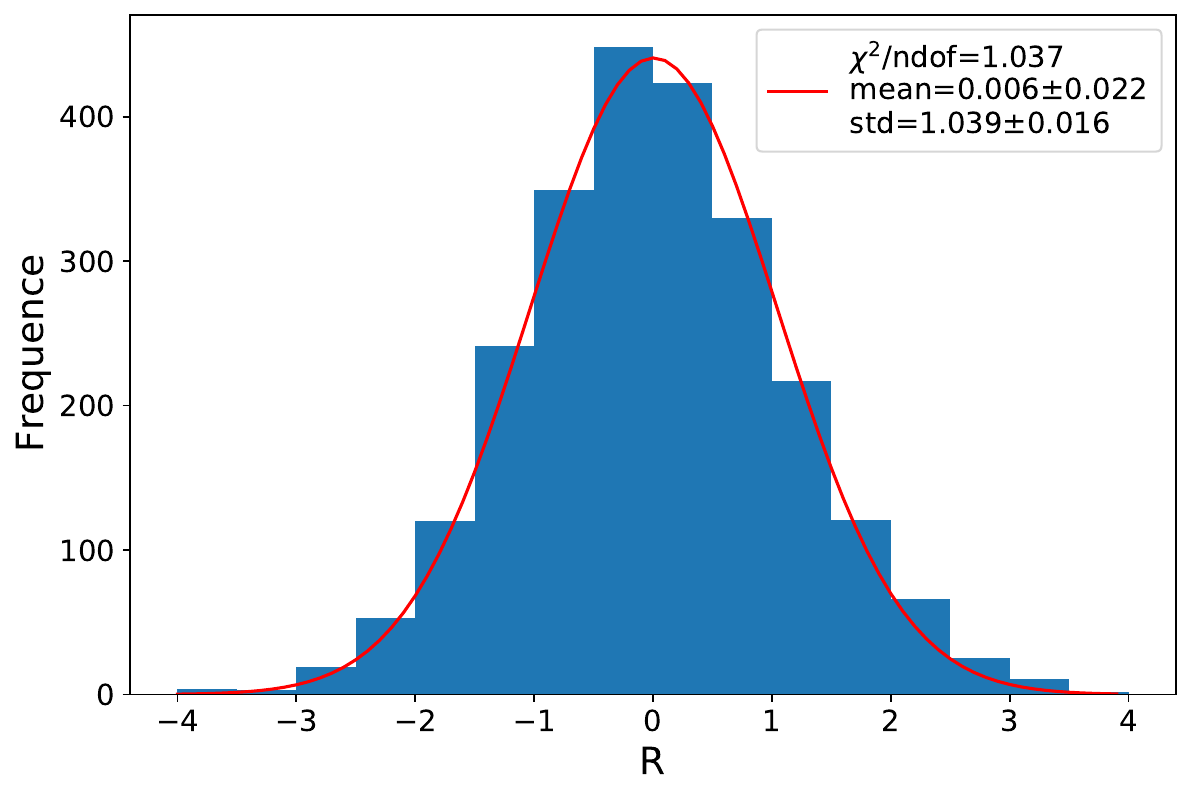}
	\caption{Upper: The upper panel is the ``super-long'' curve synthesized with several actual GRBs observed by {\it Fermi}/GBM (blue line), then reproduced by the GRB simulator and detected by GRDs (orange line). GRD\#5, \#6, \#13, \#14, \#15, \#16, \#22, \#23 and \#24 were selected . The bottom panel is the residual of theoretical curve and detected light curve. 
	Bottom: The distribution of the residual. The red line is the result of Gaussian fitting.
	}
	\label{super_long_GRB}
\end{figure}

Since the start time of the simulated GRB generated by the GRB simulator is artificially set to random, the Discrete Cross Correlation Function (DCCF) method \cite{fenimore1995gamma}\cite{link1993prevalent}\cite{band1997gamma} is applied to align the detected light curves with the theoretical light curves input to the GRB simulator on the time axis. 
Additionally, we performed a chi-square fit between the detected light curves and the theoretical curves (see eq. (\ref{theoretical_curve})) as well. 
The detected light curve and theoretical FRED curve are shown in the upper of Fig.~\ref{A-fred}, and the residual distribution of these two curves is shown in the bottom of Fig.~\ref{A-fred}.
Since the statistical fluctuations of the counts in each bin approximately follow the Poisson distribution, and we define the residual: 
    \begin{equation}
        R=(C_{\rm detected}-C_{\rm theoretical})/{\sqrt{C_{\rm theoretical}}}
    \end{equation}
where $C_{\rm detected}$, $C_{\rm theoretical}$ are the detected count rates, the theoretical count rates, respectively. The mean and standard deviation fitted with Gaussian of the residual distribution equals $-0.022\pm0.052$, $0.866\pm0.040$ (with 1.019 ``reduce-$\upchi^2$''), respectively. 
The detected super-long light curve are shown in the upper of Fig.~\ref{super_long_GRB}, and the residual distribution of the two curves is shown in the bottom of Fig.~\ref{super_long_GRB}.
The mean and standard deviation of the distribution equals $-0.006\pm0.022$, $1.039\pm0.016$ (with 1.037 ``reduce-$\upchi^2$''), respectively. 

\begin{figure}[htb]
	\centering
	\includegraphics[scale=0.4]{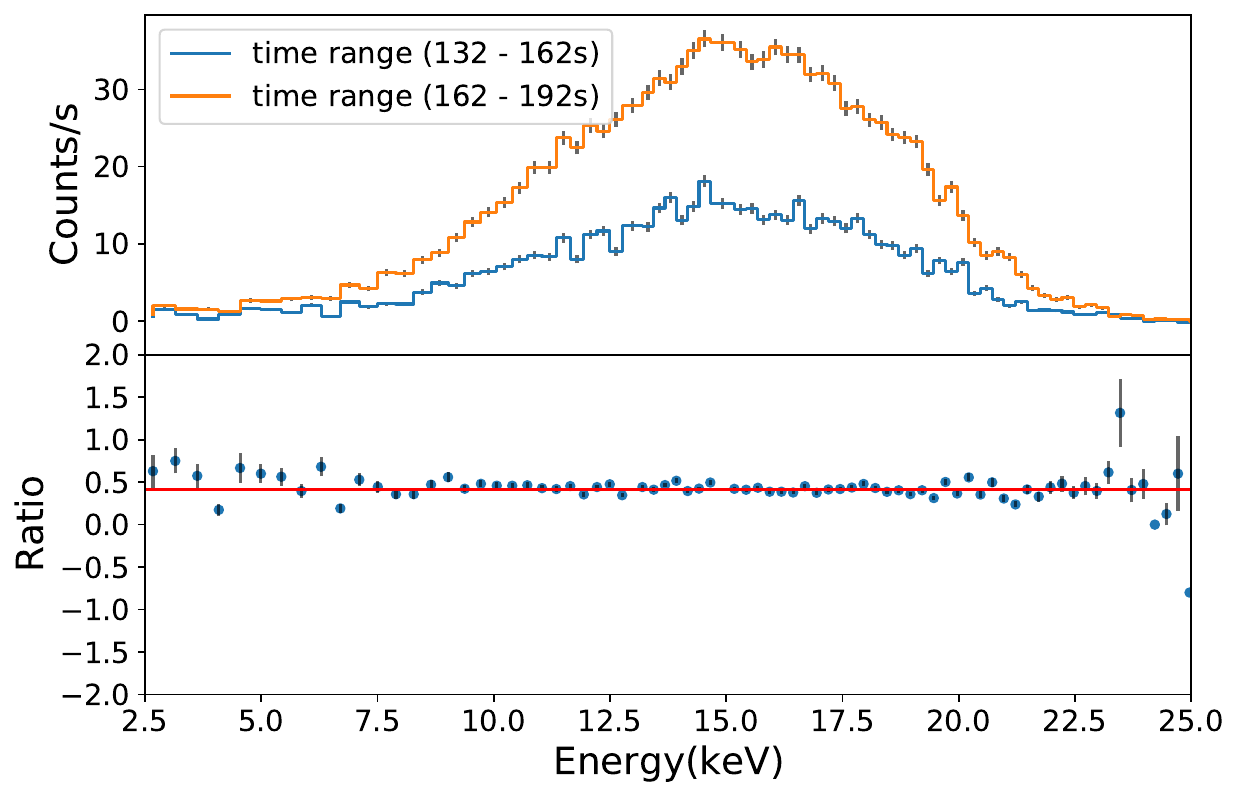}
	\caption{The shape of the energy spectrum detected by GRD in a different time range of super-long GBR (in Fig.~\ref{super_long_GRB}) is consistent which illustrates the good energy spectrum stability of the device over time.}
	\label{energy}
\end{figure}

Moreover, the stability of photon energy spectrum over time is studied by selecting two different adjacent time intervals in the super-long GRB light curve illustrated in the upper of Fig.~\ref{super_long_GRB}.
Each time interval has a 30 seconds in duration, and the total counts in the two intervals has a ratio of 1 to 2.
The energy spectra detected by GRDs in different time intervals are shown in the upper panel of Fig.~\ref{energy}, while the ratio of the two spectra for each energy bin is shown in the bottom panel of Fig.~\ref{energy}. It can be found that the photon energy spectrum is consistent in different time intervals, which proved the excellent energy stability of this GRB simulator we designed.

\section{Conclusion and discussion}\label{Conclusion}
In this work, a new GRB simulator based on grid controlled X-ray tube and direct digital synthesis technical is designed and used to verify the performance of the GECAM/GRD and the in-flight trigger and localization software of GECAM to different kinds of GRBs during the ground tests before the launch. Both the relationship of output count rates to different grid electrode voltage and the time response of the GRB simulator were investigated. The Characteristic of the GRB simulator is shown in Table~\ref{device item}, which meet the requirement of the GRB simulation experiment for GECAM.

\begin{table}[htb]
\centering
\caption{Characteristic of the GRB generator}
\label{device item}
\begin{tabular}{ccc}
\hline
Item &
  Value &
  Notes \\ \hline
Energy range &
  3 - 20 keV &
  \begin{tabular}[c]{@{}c@{}}Can be extended by applying \\ higher voltage\end{tabular} \\ \hline
Beam angle &
  About 20 \degree & -
   \\ \hline
\begin{tabular}[c]{@{}c@{}}Minimum \\ time variable scale\end{tabular} &
  About 0.9 $\upmu$s &
  Rough estimated (upper limit) \\ \hline
\begin{tabular}[c]{@{}c@{}}Maximum \\ duration time\end{tabular} &
  \begin{tabular}[c]{@{}c@{}}Theoretically \\ infinite\end{tabular} &
  \begin{tabular}[c]{@{}c@{}}The duration time divided by \\ the time binned of DDS module \\ is less than or equal to \\ 16384 (size of ROM)\end{tabular} \\ \hline
Output count rates &
  About 0 to 1 Mcps &
  \begin{tabular}[c]{@{}c@{}}Can be extended by \\ increasing filament current\end{tabular} \\ \hline
Polarization &
  - &
  \begin{tabular}[c]{@{}c@{}}Polarized radiation \\ according to \cite{muleri2017calibrating}\end{tabular} \\ \hline
\end{tabular}
\end{table}

Several GRBs generated through this GRB simulator were used to test the GECAM satellites in the configuration shown in Fig.~\ref{GECAM_exper}. 
The results show that the detected light curves are in good agreement with the input theoretical curves within the statistical uncertainty, even when the time scale of FRED is very short (with 1.6 ms time binned), demonstrating the ability of GECAM in GRB timing studies\cite{golkhou2014uncovering}\cite{kocevski2007pulse} and in the triangulating GRBs by using the time difference between the light curve detected by different satellites \cite{pal2013interplanetary}\cite{hurley2011interplanetary}\cite{hurley1999theulysses}. 
In particular, we note that the residual uncertainty is relatively large at some point with the highest brightness (i.e. the observed counts is less than the theoretical count), which probably caused by the effect of detector deadtime (approximately 4 $\upmu$s per event).

Furthermore, a 100-meter X-ray space telescope testing facility \cite{zhao2019Wolter} has been built by IHEP in Beijing, and the portable GRB simulator can be installed at one end of the facility through the standard KF40 vacuum flange to obtain quasi-parallel light at the other end of the facility without the requirement of a collimator. 

Considering the limitation of heat dissipation, the maximum voltage that could be applied to the X-ray tube we designed is about 20 kV, which means that this GRB simulator can only produce X-rays below about 20 keV (much softer than normal GRBs). 
In our next work, water cooling will be added to the X-ray tube to increase the rated voltage, achieving the modulate of higher X-ray energy up to 50 keV or more, which will allow us to mimic GRBs with normal spectral hardness.

\section*{Acknowledgements}
This research is supported by the Strategic Priority Program on Space Science, the Chinese Academy of Sciences, Grant No. XDA15020501, No. XDA1531010301 and No. XDA15360102. We are also grateful to Shanghai KEYWAY ELECTRON Technology Co., Ltd for their assistance in the processing and design of X-ray tubes.

\section*{Conflict of interest}

The authors declare that they have no conflict of interest.

\bibliography{reference}

\end{document}